\documentclass[aps,pra,preprint,superscriptaddress,longbibliography]{revtex4-2}

\usepackage{amsmath}
\usepackage{amsthm}
\usepackage{amsfonts}
\usepackage{amssymb}
\usepackage[dvipsnames]{xcolor}
\usepackage{graphicx}
\usepackage{tikz}
\usepackage{color}
\usepackage[pdftex]{hyperref} 
\usepackage{longtable}
\usepackage{array}
\usepackage{dsfont}
\usepackage{bm}
\usepackage{physics}
\usepackage{algorithm}
\usepackage{algpseudocode}

\renewcommand{\vec}[1]{\ensuremath{\mathbf{#1}}}
\newcommand{\hatt}[1]{\ensuremath{\mathbf{ \hat{#1} }}}

\begin{document}

    \title{Towards polarization steganography}

    \author{Valeria Tena--Piñon}
    \affiliation{Photonics and Mathematical Optics Group, Tecnologico de Monterrey, Monterrey 64849, Mexico}

    \author{Atefeh Akbarpour}
    \affiliation{Photonics and Mathematical Optics Group, Tecnologico de Monterrey, Monterrey 64849, Mexico}

    \author{Przemyslaw Litwin}
    \affiliation{Wrocław University of Science and Technology, Department of Optics and Photonics, Wybrzeże Wyspiańskiego 27, 50-370 Wrocław, Poland}

    \author{Adad Yepiz}
    \affiliation{Photonics and Mathematical Optics Group, Tecnologico de Monterrey, Monterrey 64849, Mexico}

    \author{Fernando Torres--Leal}
    \affiliation{Department of Physics and Astronomy, University of Rochester, Rochester, New York 14627, USA}
    \affiliation{Photonics and Mathematical Optics Group, Tecnologico de Monterrey, Monterrey 64849, Mexico}

    \author{\\Raul I. Hernandez--Aranda}
    \affiliation{Photonics and Mathematical Optics Group, Tecnologico de Monterrey, Monterrey 64849, Mexico}

    \author{Mateusz Szatkowski}
    \affiliation{Wrocław University of Science and Technology, Department of Optics and Photonics, Wybrzeże Wyspiańskiego 27, 50-370 Wrocław, Poland}

    \author{Blas M. Rodriguez--Lara}
    \affiliation{Universidad Polit\'ecnica Metropolitana de Hidalgo, Tolcayuca, Hidalgo 43860, Mexico}

    \author{Benjamin Perez--Garcia}
    \email[Corresponding author: ]{b.pegar@tec.mx}
    \affiliation{Photonics and Mathematical Optics Group, Tecnologico de Monterrey, Monterrey 64849, Mexico}

    \date{\today}

    \begin{abstract}
        We propose and experimentally demonstrate a polarization--based steganographic scheme using partially polarized vector beams. 
        In our approach, the spatially dependent polarization structure of the optical field serves as the carrier through which the hidden information can be retrieved.
        By engineering a vector beam whose polarization states populate a prescribed region of the Poincar\'e sphere, specifically, the equatorial disk, we establish a nontrivial mapping between transverse spatial coordinates and polarization states. 
        Information retrieval is achieved by applying a spatial mask derived from a parametric curve defined within this region of the Poincar\'e sphere, followed by spatially resolved polarization analysis. 
        We demonstrate the selective reconstruction of various parametric shapes, including polygonal and smooth curves, confirming that the hidden patterns are retrieved through the combined use of spatial filtering and polarization--domain mapping. 
        Our results establish partially polarized vector beams as a flexible and experimentally accessible platform for polarization--based information hiding.
    \end{abstract}

\maketitle
\newpage	


\section{Introduction}

Polarization is a fundamental property of electromagnetic waves that describes the temporal evolution of the electric field vector at a given point in space \cite{Goldstein2011}. 
In the paraxial regime, polarization is commonly characterized by the relative amplitudes and phases of the transverse field components, allowing for a complete description in terms of Jones matrices and Stokes parameters \cite{Chipman2018}.
Light beams that feature space-dependent polarization states are known as vector beams \cite{Zhan2009}. 
These optical fields have attracted considerable attention in recent years, spanning both fundamental studies and applied research. 
Representative examples include their use in the characterization of quantum channels \cite{Ndagano2017a}, as well as in the implementation of quantum algorithms \cite{Zhang2012, Sephton2019} and cryptographic protocols \cite{Chen2016, Ndagano2017b, Sit2017}. 
Vector beams have also found applications in industrial contexts, for instance in microdrilling and other laser--based material processing tasks \cite{Matsusaka2018}. 
From a theoretical standpoint, vector light has recently been explored in the context of optical skyrmions \cite{Ye2024}.
M\"obius strips and other nontrivial topological structures have likewise been employed to engineer and manipulate the polarization structure of light \cite{Bauer2015, Galvez2017, Larocque2020}.

Seminal works have investigated vector light in the partially polarized regime \cite{Martinez2009, Zhan2014}; however, the majority of reported applications focus on beams with a high degree of polarization \cite{Rosales-Guzman2018}. 
Here, we present a proof of principle experiment in which we explore the use of partially polarized vector beams as a platform for information hiding. 
The encoded information is retrieved by applying an appropriate spatial filter followed by a spatially resolved polarization mapping to the Poincar\'e sphere.

\begin{figure}[htp!]
  \centering
  \includegraphics[width=80mm]{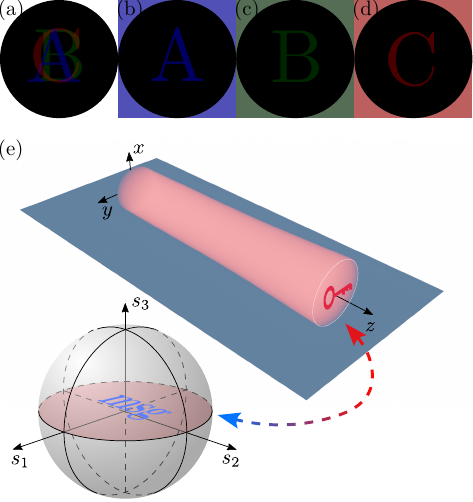}
  \caption{(a)--(d) RGB Steganography. Three letters are encoded using distinct colors: A (blue), B (green), and C (red). 
  (d) Spatial overlap of the three letters, which makes it difficult to isolate the individual information channels; however, applying a matching color filter reveals the corresponding letter. 
  (e) Concept of polarization steganography. A suitable spatial filter (red key in the diagram) applied in the transverse $(x,y)$ plane of the field reveals the information embedded (“msg”) within the equatorial disk of the Poincaré sphere.}
  \label{fig:fig01}
\end{figure}

\section{Concept and theory}

Steganography concerns the concealment of information within another medium \cite{Kahn1996}. 
To understand this, let us present the following example.
Consider an image composed of three overlapping letters, denoted A, B, and C. 
Each letter is encoded using a distinct color, for instance blue, green, and red, respectively. 
Owing to the spatial overlap of the three letters, it is difficult to isolate the individual information channels; however, applying a matching color filter reveals the corresponding letter (see Fig.\ \ref{fig:fig01}(a)--(d)).

Optical implementations of steganography have been explored using different degrees of freedom of light. 
For instance, digital holography has been employed to hide a secret image within a cover image, with subsequent retrieval achieved through optical processing \cite{Hamam2010}. 
Other approaches include ghost steganography based on computational ghost imaging \cite{Liu2020}, as well as metasurface-based optical steganography, where hidden information is encoded in the optical response of nanostructured surfaces \cite{Song2021}.

In our approach, the partially polarized optical field acts as the carrier, while the concealed information is encoded in the nontrivial mapping between transverse spatial coordinates and local polarization states. 
Information retrieval is achieved through a carrier-dependent spatial mask that selectively samples those beam coordinates whose polarization states satisfy a prescribed parametric curve on the Poincar\'e sphere.
In particular, we employ the equatorial disk of the Poincar\'e sphere for this purpose. 
Our first task is to engineer an optical field whose polarization states populate this region in a controlled manner. 
One possible route to achieve this goal is to illuminate a $q$--plate \cite{Marrucci2006} with a quasi--monochromatic extended source.

A $q$--plate is a patterned birefringent optical element that enables spin--orbit coupling \cite{Marrucci2006}. 
Its fast axis orientation varies azimuthally in the transverse plane, resulting in a polarization-dependent modulation. 
As a consequence, an incident beam with linear homogeneous polarization emerges with a spatially structured polarization state determined by the parameter $q$. 
Mathematically, the action of a $q$--plate immediatly after the element, for a coherent and linearly polarized input field $\vec{E}_\text{in}(\vec{r}) = E_{\text{in}}(\vec{r})\hatt{x}$, is of the form
\begin{align}
	\vec{E}_\text{out}(\vec{r}) = E_{\text{in}}(\vec{r}) \exp(i2q\theta)\hatt{e}_L + E_{\text{in}}(\vec{r}) \exp(-i2q\theta)\hatt{e}_R,
\end{align}
where $\vec{r}$ is the transverse position vector, $\theta$ is the azimuthal angle, and $\hatt{e}_L$ and $\hatt{e}_R$ represent the left-- and right--handed circular polarization unit vectors.

\begin{figure}[htp!]
  \centering
  \includegraphics[width=80mm]{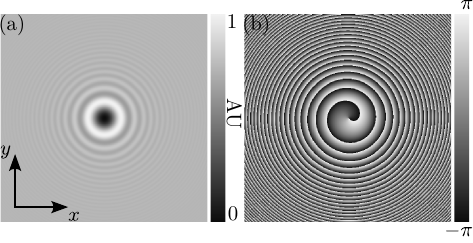}
  \caption{(a) Far field intensity and (b) transverse phase, of a spherical source realization after propagation through a $q$--plate, computed from Eq.\ \ref{eq:oam} with parameters $\ell=1$, $f=15$ cm, $z=50$ cm, and $\lambda = 625$ nm. The numerical window has a side length of 2 mm.}
  \label{fig:fig02}
\end{figure}

To model the extended source, we adopt the beam--wander framework \cite{Gbur2016}. 
Within this model, each centroid point of the source is regarded as a spherical emitter. 
The position vector $\vec{r}_0$ of each point source is treated as a random variable drawn from a uniform circular probability distribution. 
The field emitted from each point propagates over a distance $z$, subsequently passes through a $q$--plate, and the resulting field is propagated to the far field by means of a lens, thus computing its Fourier transform \cite{Goodman1996}.
Finally, we compute the average over $\vec{r}_0$.

A single scalar realization of the field described above, prior to ensemble averaging, can be written as \cite{Olver2010}
\begin{align} \nonumber \label{eq:oam}
	U_\ell(\vec{r},z) = &A_0 \sqrt{\frac{z}{2}} \frac{k}{f} \left(\frac{2\pi}{k}\right)^{3/2} \exp(i3\pi|\ell|/4) \times \\ \nonumber
	& \hspace{-30pt}\exp\left(i\frac{k}{2z}\abs{\vec{r}_0}^2\right) \abs{\vec{r}-\frac{f}{z}\vec{r}_0 } \exp\left(-i\frac{kz}{4f^2}\abs{\vec{r}-\frac{f}{z}\vec{r}_0}^2\right) \times \\ \nonumber
	& \hspace{-30pt}\left[J_{\frac{|\ell|+1}{2}}\left(\frac{kz}{4f^2}\abs{\vec{r}-\frac{f}{z}\vec{r}_0}^2\right) + i J_{\frac{|\ell|-1}{2}}\left(\frac{kz}{4f^2}\abs{\vec{r}-\frac{f}{z}\vec{r}_0}^2\right)\right]\times \\
	& \hspace{-30pt}\exp[i\ell\:\text{atan}\left(\frac{y-\frac{f}{z}y_0}{x-\frac{f}{z}x_0} \right)].
\end{align}
Here, $J_\nu(\cdot)$ represents the Bessel function of the first type of order $\nu$, $\vec{r}=x\hatt{x}+y\hatt{y}$ denotes the transverse position vector at the observation plane, $z$ is the propagation distance from the source to the $q$-plate, and $\ell=2q$ is the topological charge. 
The wavenumber is given by $k=2\pi/\lambda$, $f$ denotes the focal length of the lens used to reach the far field, and $\vec{r}_0=x_0\hatt{x}+y_0\hatt{y}$ represents the transverse position of the spherical source at $z=0$.
Figure \ref{fig:fig02} shows the corresponding transverse intensity and phase for the case $\ell=1$.

\begin{figure}[htp!]
  \centering
  \includegraphics[width=80mm]{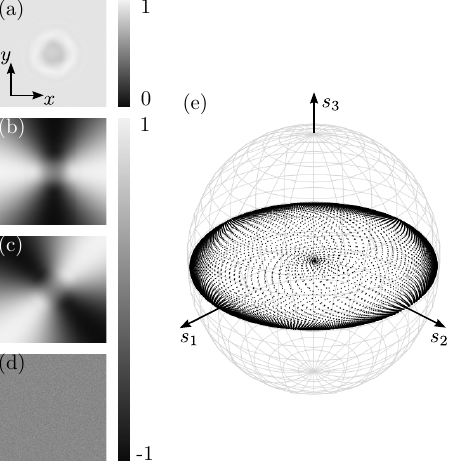}
  \caption{(a)--(d) Theoretical Stokes parameters $S_0, S_1, S_2, S_3$, respectively, and (e) their corresponding map to the Poincar\'e sphere. We used the following parameters:  $\ell=1$, $a=3.75\times 10^{-4}$ m, $\lambda = 625$ nm, $f = 15$ cm, and $z = 24$ cm. The numerical window has a side length of 2 mm.}
  \label{fig:fig03}
\end{figure}

The local polarization matrix describing our partially polarized field is given by \cite{Gori1998, Martinez2009}
\begin{align}\label{eq:bcp}
\mathbb{J}^\ell(\vec{r},z) =
\begin{bmatrix} \langle U_\ell(\vec{r},z) U^*_\ell(\vec{r},z) \rangle & \langle U_\ell(\vec{r},z) U^*_{-\ell}(\vec{r},z) \rangle \\ \langle U_{-\ell}(\vec{r},z) U^*_\ell(\vec{r},z) \rangle & \langle U_{-\ell}(\vec{r},z) U^*_{-\ell}(\vec{r},z) \rangle \end{bmatrix}.
\end{align}
The angular brackets denote an ensemble average over the source realizations, defined as
\begin{align}
\langle h(\vec{r},\vec{r}_0) g(\vec{r},\vec{r}_0) \rangle = \iint_{\Omega} f(\vec{r}_0) g(\vec{r},\vec{r}_0) h(\vec{r},\vec{r}_0) d^2\vec{r}_0,
\end{align}
where $f(\vec{r}_0)$ is the probability density function associated with the transverse position $\vec{r}_0$ of the spherical emitters at $z=0$, and $\Omega$ denotes its support.
Here, we consider a uniform circular distribution of radius $a$.

From the elements in Eq.\ \ref{eq:bcp}, we can compute the Stokes parameters via \cite{Martinez2009, Zhan2014}
\begin{align}
	S_0(\vec{r},z) &= J^\ell_{11}(\vec{r},z) + J^\ell_{22}(\vec{r},z),\\
	S_1(\vec{r},z) &= 2\text{Re}\{J^\ell_{12}(\vec{r},z)\},\\
	S_2(\vec{r},z) &= 2\text{Im}\{J^\ell_{12}(\vec{r},z)\},\\
	S_3(\vec{r},z) &= J^\ell_{11}(\vec{r},z) - J^\ell_{22}(\vec{r},z).
\end{align}
The degree of polarization may be then obtained either from the Stokes parameters or, equivalently, directly from the local polarization matrix as
\begin{align}
	p(\vec{r},z) = \sqrt{1 - 4\frac{\det[\mathbb{J}^\ell(\vec{r},z)]}{(\Tr[\mathbb{J}^\ell(\vec{r},z)])^2}},
\end{align}
where $0 \le p(\vec{r},z) \le 1$.

Given the ingredients just defined, let us plug Eq.\ \ref{eq:oam}, into Eq.\ \ref{eq:bcp}.  
To further analyze the model, we evaluate the ensemble averages numerically, owing to the complexity of obtaining closed--form analytical expressions for these integrals.
From the resulting polarization matrix, we compute the Stokes parameters and construct the corresponding mapping onto the Poincar\'e sphere. 
The theoretical results are shown in Fig.\ \ref{fig:fig03}, where we observe that the equatorial disk is populated by a collection of spots, as required by our scheme.

\begin{figure}[htp!]
  \centering
  \includegraphics[width=80mm]{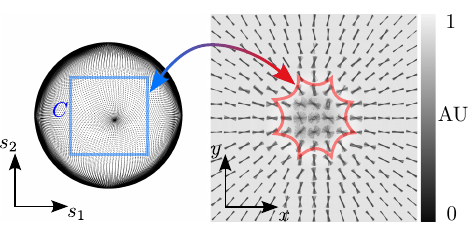}
  \caption{Illustration of the spatial--polarization mapping. A suitable spatial filter applied in the $(x,y)$ plane translates into a predefined shape within the equatorial disk of the Poincar\'e sphere (left--hand panel). The red trajectory in the $(x,y)$ plane maps onto the blue curve in the $(s_1,s_2)$ plane. The right--hand panel depicts $S_0$, together with representative electric field trajectories, consistent with the corresponding polarization states.}
  \label{fig:fig04}
\end{figure}

In our scheme we generate a vector beam with a spatially dependent degree of polarization.  
As discussed above, we specifically consider a vector beam whose mapping onto the Poincar\'e sphere fills the equatorial disk.
Now, we consider a parametric curve 
\begin{align}
	C : \vec{C}(t) = C_1(t)\hatt{s}_1 + C_2(t)\hatt{s}_2 + C_3(t)\hatt{s}_3, \text{ } 0\le t \le T,
\end{align}
which lies entirely within the equatorial disk of the Poincar\'e sphere, that is, $C_3(t)=0$. 
The objective is to identify the transverse spatial points $(x,y)$ of the field whose associated polarization states coincide with the parametric curve $C$ on the Poincar\'e sphere (see Fig.\ \ref{fig:fig04}). 
This task can be accomplished by numerically inverting the Stokes parameters corresponding to the prescribed curve $C$ (see Algorithm \ref{alg:01}).
We note that this procedure depends on the spatial distribution of polarization states across the field.

\begin{algorithm}[H]
\caption{Spatial mask construction}
\begin{algorithmic}[1]

\State \textbf{Input:} Stokes distributions $(S_0,S_1,S_2,S_3)$, curve $(C_1,C_2,C_3)$, tolerance $\varepsilon$
\State \textbf{Output:} Spatial mask $M$

\State Compute normalized Stokes parameters
\[
s_1 = S_1/S_0, \qquad s_2 = S_2/S_0, \qquad s_3 = S_3/S_0
\]

\For{each pixel $(i,j)$}

	\For{each index $(t)$}
		\If {$(s_1(i,j)-C_1(t))^2+(s_2(i,j)-C_2(t))^2+(s_3(i,j)-C_3(t))^2 < \varepsilon^2$}
			\State $M(i,j) \leftarrow 1$
		\Else
			\State $M(i,j) \leftarrow 0$
		\EndIf
	\EndFor
\EndFor

\end{algorithmic}
\label{alg:01}
\end{algorithm}

\section{Experiment and results}

\begin{figure}[htp!]
  \centering
  \includegraphics[width=80mm]{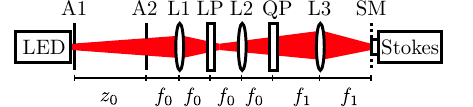}
  \caption{Schematic layout of experimental setup.  LED, extended source; A1--A2, apertures; L1--L3, lenses; LP, linear polarizer; QP, $q$--plate; SM, spatial mask; Stokes, spatially resolved Stokes polarimetry.}
  \label{fig:fig05}
\end{figure}

Figure \ref{fig:fig05} illustrates our experimental setup for generating the partially polarized vector beam \cite{Zeng2020}. 
An LED source (Thorlabs M625F2, central $\lambda = 625$ nm), together with a variable aperture (A1), is propagated over a distance of $z_0=24$ cm to a second aperture (A2). 
The field at A2 passed through a horizontal linear polarizer and is then imaged onto a $q$--plate with $q=1/2$ by a pair of lenses, each with focal length $f_0=12.5$ cm. 
A subsequent lens with focal length $f_1=15$ cm, placed 15 cm after the $q$--plate, is used to form the far field at its back focal plane. 
A CCD camera (Thorlabs DCU224M) positioned at this plane is employed to record the resulting field distribution.
The Stokes parameters were measured using standard polarization optics, consisting of a linear polarizer and a quarter--wave plate \cite{Goldstein2011}.

\begin{figure}[htp!]
  \centering
  \includegraphics[width=80mm]{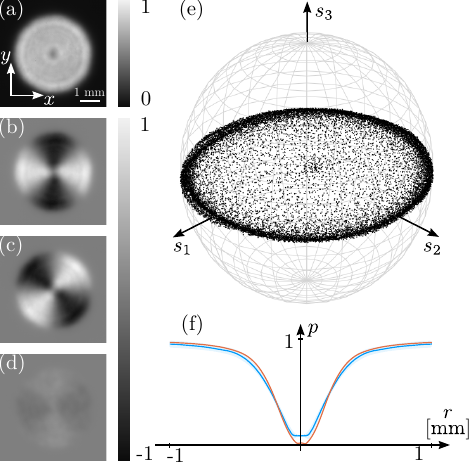}
  \caption{(a)--(d) Experimental Stokes parameters $S_0, S_1, S_2, S_3$, respectively, and (e) their corresponding map to the Poincar\'e sphere. (f) Shows the degree of polarization as a function of the radial coordinate (orange theory, blue experiment). We set the radius of A1 to $3.75\times 10^{-4}$ m.}
  \label{fig:fig06}
\end{figure}

Figures \ref{fig:fig06}(a)--(d) display the experimentally obtained Stokes parameters for the case $\mathbb{J}^{\ell=1}(\vec{r},z)$. 
Figure \ref{fig:fig06}(e) shows the corresponding mapping of these parameters onto the Poincar\'e sphere. 
From these results, we experimentally confirm that the generated field populates the equatorial disk. 
Figure \ref{fig:fig06}(f) presents the radial dependence of the degree of polarization. 
The orange curve corresponds to the theoretical prediction, while the blue curve represents the experimental measurement. 
The light blue shaded region indicates the experimental uncertainty associated with the measurements.

We implemented the spatial masks --tested both physically and digitally-- computed from prescribed parametric equations. 
The physical spatial masks were generated using high--resolution printing with an imagesetter (Agfa AVANTRA 44), operating at 2400 Dots Per Inch (DPIs), corresponding to a pixel size of approximately 10 $\mu$m.
The digital spatial masks were achieved by applying the corresponding filtering directly to the spatially resolved measured Stokes parameters --in the form of numerical matrices-- for the field shown in Fig.\ \ref{fig:fig06}.

For the physical spatial masks implementation, we realized two cases: a square frame and an astroid, shown in Figs.\ \ref{fig:fig07}(a) and (b), respectively.
For the digital spatial masks implementation, we realized three additional cases: a cardioid, a heart--shaped curve, and a Lissajous figure, shown in Figs.\ \ref{fig:fig07}(c)--(e), respectively.
All parametric definitions are listed in Table \ref{table:parametric}.
Figure \ref{fig:fig07}(f) serves as a reference when no spatial mask is present.
The corresponding spatial masks are shown as insets in Fig.\ \ref{fig:fig07}.
The retrieved results for all cases (physical and digital) closely reproduce the intended two--dimensional curves mapped onto the equatorial disk of the Poincar\'e sphere.

\begin{table}[htp!]
	\centering
	\caption{Definition of the parametric equations used to construct the spatial masks in this work \cite{Lockwood1967}.}
	\label{table:parametric}
	\begin{tabular}{ll}
		{\bf Name} & {\bf Parametric equations} \\
		\hline 
		\hline
		Frame 			& $C_1(t) = \frac{A}{2} \text{sgn}(\cos t) \min\left(1, \frac{|\cos t|}{|\sin t|}\right) $ \\
						& $C_2(t) = \frac{A}{2} \text{sgn}(\cos t) \min\left(1, \frac{|\sin t|}{|\cos t|}\right)$ \\
		\hline
		Astroid 	  	& $C_1(t) = A \cos^3 t$ \\
						& $C_2(t) = A \sin^3 t$ \\
		\hline
		Cardioid		& $C_1(t) = 2A (1-\cos t)\cos t$ \\
						& $C_2(t) = 2A (1-\cos t)\sin t$ \\
		\hline
		Heart			& $C_1(t) = 16A \sin^3 t$ \\
						& $C_2(t) = A (13\cos t - 5\cos 2t - 2\cos 3t - \cos 4t)$ \\
		\hline
		Lissajous		& $C_1(t) = A_1 \cos (\omega_1 t)$ \\
						& $C_2(t) = A_1 \cos (\omega_2 t + \delta)$ \\   
	\end{tabular}
\end{table}

\section{Discussion}

\begin{figure*}[h]
  \centering
  \includegraphics[width=80mm]{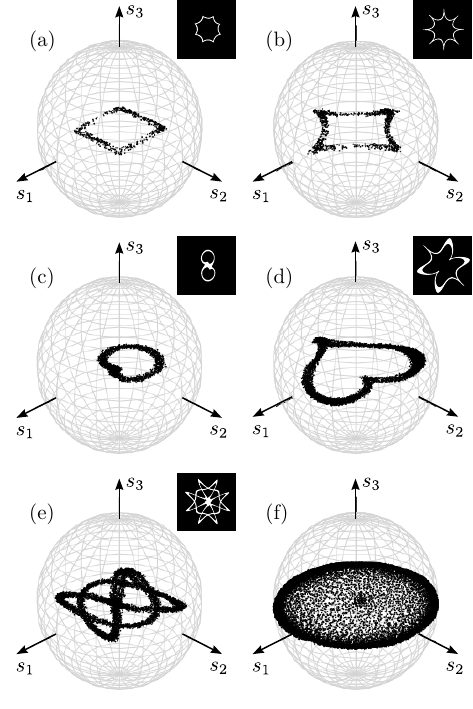}
  \caption{Experimentally filtered polarization states mapped onto the Poincar\'e sphere for different designed curves. Physical spatial masks: (a) Square frame ($A=0.75$) and (b) astroid ($A=0.75$).  Digital spatial mask: (c) cardioid ($A=0.15$), (d) heart-shaped curve ($A=0.05$), and (e) Lissajous curve ($A_1=A_2=0.6$, $\omega_1/\omega_2 = 2/3$, $\delta=\pi/4$).  (f) depicts the reference case with no spatial mask applied. The insets show the spatial masks (transmittance filters) for each curve with side length of 1 mm. In all cases, we set the radius of A1 to $3.75\times 10^{-4}$ m.}
  \label{fig:fig07}
\end{figure*}

The results presented in Fig.\ \ref{fig:fig07} demonstrate the effectiveness of our approach for encoding and retrieving information using partially polarized vector beams. 
In the absence of a spatial mask (Fig.\ \ref{fig:fig07}(f)), the mapping of the measured Stokes parameters onto the Poincar\'e sphere populates the equatorial disk, as expected from the engineered spatial distribution of polarization states. 
In this case, no specific structure can be identified, and the encoded information remains concealed within the polarization degree of freedom.
When a spatial mask derived from a prescribed parametric curve is applied (Figs.\ \ref{fig:fig07}(a)--(e)), a clear structural signature emerges in the polarization-domain representation. 
In each case, only those transverse spatial points whose polarization states coincide with the target curve on the Poincar\'e sphere are transmitted, resulting in a selective reconstruction of the intended shape. 
This behavior confirms that the information retrieval process is governed by the joint action of spatial filtering and the underlying polarization-state distribution.

The five representative cases considered: a square frame, an astroid, a cardioid, a heart-shaped curve, and a Lissajous figure, span both simple polygonal geometries and smooth, highly nontrivial parametric curves. 
The successful reconstruction of all these shapes highlights the flexibility of the proposed scheme and its independence from a particular class of curves. 
The method relies on the existence of a mapping between transverse spatial coordinates and polarization states within a region in the Poincar\'e sphere.
We emphasize that this mapping depends critically on the specific polarization state distribution of the optical field. 
Consequently, the retrieval mask is not a universal representation of the target curve, but a carrier-dependent key derived from the particular spatial polarization structure of the beam. 
As a proof of concept, we employed cylindrical vector beams generated by a q-plate, which produce a symmetric polarization state distribution. 
More generally, the polarization distibution can be engineered to follow more intricate patterns beyond the symmetric case considered here, enabling more complex carrier structures \cite{GonzalezAceves2025}.

From a steganographic perspective, our results show that polarization can serve as an effective channel for hidden information, while the spatial intensity distribution alone does not directly reveal the encoded pattern. 
In the present scheme, retrieval is achieved by combining spatial filtering with spatially resolved polarization mapping onto the Poincar\'e sphere, thereby selecting the subset of beam coordinates associated with the target parametric curve.

\section{Conclusion}

In this work, we have introduced and experimentally demonstrated a polarization-based steganographic scheme that exploits the spatially varying degree of polarization of partially polarized vector beams. 
By engineering an optical field whose polarization states populate the equatorial disk of the Poincar\'e sphere, we established a nontrivial mapping between transverse spatial coordinates and polarization states that serves as the foundation for information hiding.
We showed that information can be selectively retrieved by applying spatial masks derived from prescribed parametric curves defined within the Poincar\'e sphere. 
The experimental reconstruction of a variety of geometrical shapes, ranging from simple polygonal frames to smooth and highly nontrivial curves, confirms the versatility of the proposed approach. 
Although the examples considered here exhibit a certain degree of symmetry, the method is not restricted to symmetric curves $C$ and can be extended to more general, non--symmetric shapes.
In the absence of the appropriate spatial filter and polarization-domain analysis, the hidden pattern remains inaccessible, highlighting the selective retrieval enabled by the scheme.

Our theoretical model, based on the beam wander description of an extended source and ensemble averaging, accurately captures the essential features of the generated partially polarized vector field and is in good agreement with experimental observations. 
The use of standard optical components, including an LED source, $q$--plate, and conventional polarimetric elements, underscores the experimental accessibility of our proof of principle proposal.

Beyond steganography, the concepts introduced here provide a general framework for exploiting partial polarization and structured polarization state distributions as additional control parameters in structured light.


\section*{Funding}
This work was partially funded by Narodowe Centrum Nauki (2022/45/B/ST7/01234). 

\section*{Acknowledgments}
B.P.--G. acknowledge fruitful discussions with Mitch Cox, Juan P. Rubio--Perez and Hugo A. Moreno--Rodriguez. 
B.~M.~R.~L. acknowledges support and hospitality as an affiliate visiting colleague at the Department of Physics and Astronomy, University of New Mexico.

\section*{Disclosures}
The authors declare no conflicts of interest.

\section*{Data Availability Statement}
All the data is available from the corresponding author upon reasonable request.



%

\end{document}